\documentclass[twocolumn,showpacs,preprintnumbers,amsmath,amssymb,prd]{revtex4}

\usepackage{graphicx}% Include figure files
\usepackage{bm}% bold math

\begin{document}

\newcommand{\sect}[1]{Sec.~\ref{sec:#1}}
\newcommand{\eqn}[1]{Eq.~(\ref{eq:#1})}
\newcommand{\etal}{\mbox{\textit{et al.}}}
\newcommand{\fig}[1]{Fig.~\ref{fig:#1}}

%  SLAC pub header required information
\preprint{SLAC-PUB-9191}

\title{Large bulk matter search for fractional charge particles}
%  SLAC pub header required information
\thanks{Work supported by Department of Energy contract DE-AC03-76SF00515.}

\author{
  Irwin T. Lee, Sewan Fan, Valerie Halyo, Eric R. Lee, \\
  Peter C. Kim, Martin L. Perl and Howard Rogers
}

\affiliation{
  Stanford Linear Accelerator Center, Stanford University, \\
  Stanford, California 94309
}

\author{
  Dinesh Loomba
}
\affiliation{
  Department of Physics, University of New Mexico, \\
  Albuquerque, New Mexico 87131
}

\author{
  Klaus S. Lackner
}
\affiliation{
  Department of Earth and Environmental Engineering, Columbia University, \\
  New York, New York 10027
}

\author{
  Gordon Shaw
}
\affiliation{
  Department of Physics, University of California--Irvine, \\
  Irvine, California 92717
}

%  SLAC pub required header information
\author{Submitted to Physical Review D}

%PRD_ONLY\date{April 3, 2002}
\date{April 2, 2002}

\begin{abstract}
We have carried out the largest search for stable particles with
fractional electric charge, based on an
oil drop method that incorporates
a horizontal electric field and upward air flow.
No evidence for such particles was found, giving a 95\% confidence
level upper limit of
$1.17\times 10^{-22}$ particles per nucleon on the abundance of fractional
charge particles in silicone oil for $0.18 e \le |Q_{residual}| \le 0.82 e$.
Since this is the first use of this new method we describe the advantages and
limitations of the method.
\end {abstract}


\maketitle

\section{INTRODUCTION}
\label{sec:introduction}

We have carried out the largest search for fractional electric charge
elementary particles in bulk matter using 70.1~mg of silicone oil.  That is,
we looked for stable particles whose charge $Q$ deviates from $Ne$ where $N$
is an integer, including zero, and $e$ is the magnitude of the charge on the
electron.  No evidence for such particles was found in this amount of silicone
oil.  We used our new version \cite{loomba} of the Millikan oil drop method
containing two innovations compared to the classical method that we used in
Halyo \etal \cite{halyo}.  One innovation is that the drop charge is
obtained by observing the drop motion in a \emph{horizontal}, alternating
electric field compared to the classical use of a vertical electric
field \cite{halyo, mar, savage}.  The other innovation is the use of an upward
flow of air to reduce the vertical terminal velocity of the drop,
which enabled us to use larger drops, about
20.6~$\mu$m in diameter compared to the 10~$\mu$m drops used in our previous
experiments.

We define the residual drop charge, $Q_r = Q-N_le$ where $N_l$ is
the largest integer less than $Q/e$.  We find the 95\% confidence level
upper limit on the abundance of fractional charge particles in silicone oil
for $0.18 e \le Q_r \le 0.82 e$ is $1.17\times 10^{-22}$ particles
per nucleon.  This experiment was a follow up on our previous search in
silicone oil, Halyo \etal \cite{halyo}, based on 17.4~mg.  In that
search we found one drop with anomalous charge, but no such charge was
found in the present experiment.

In this paper we describe the experimental method and apparatus in
\sect{method_and_apparatus}.  In \sect{calibration_and_errors} we discuss
the measurement precision resulting
from the various measurement errors and the calibration methods.  The
data analysis method, including the criteria used to accept drop
charge measurements, is discussed in \sect{analysis_and_results}.
Here we pay particular
attention to the drop spacing criterion necessitated by
interactions between adjacent drops.  This is the primary limitation on
the rate at which drops can be measured and we had to acquire
considerable experience to understand this limitation.  We conclude
with \sect{conclusions}, giving our results, comparing our results with other
fractional charge searches, and discussing the applicability and
extension of this new Millikan oil drop technique to other searches.

\section{EXPERIMENTAL METHOD AND APPARATUS}
\label{sec:method_and_apparatus}

\subsection{Experimental method}

\begin{figure}
\resizebox{\columnwidth}{!}{
	\includegraphics{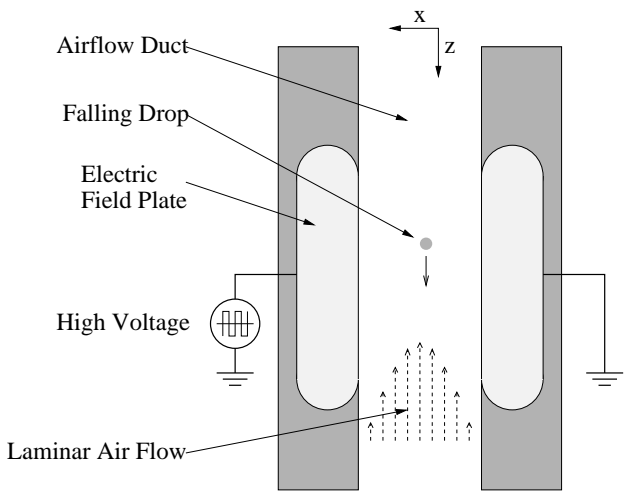}
}
\caption{Basic principles of the experimental method.}
\label{fig:principles}
\end{figure}

The principle of the experimental method is simple.  Consider a drop of radius
$r$, density $\rho$, and charge $Q$ falling in air through a horizontal
electric field of strength $E$, as shown in \fig{principles}.
Applying Stokes' law the
horizontal terminal velocity, $v_x$ is 
\begin{equation}
v_x = \frac{QE}{6\pi\eta r}
\label{eq:stokes}
\end{equation}
where $\eta$ is the viscosity of air.  Hence measuring $v_x$ gives $Q$
providing $r$ is known.  As explained in \sect{field_and_radius},
the drop radius is
determined from the $v_x$ of integer charge particles.  Note that the
measurement of $Q$ does not depend on the density of the drop and is
also independent of the gravitational force on the drop.  The electric
field alternates in the $+x$ and $-x$ direction so that the drop is
moved back and forth along the $x$~axis.  This cancels some sources of
error and allows the drop motion to be viewed in a relatively narrow
horizontal area.  Previous uses of a static horizontal electric field
in the Millikan oil drop method were in 1941 by Hopper and
Laby \cite{hopper} who measured the electron charge and by
Kunkel \cite{kunkel} in 1950 who measured the charge on dust
particles.

If the drop were falling in still air, the vertical terminal velocity
would be given by
\begin{equation}
v_{z,term} =\frac{2 r^2 \rho g}{9\eta}
\label{eq:terminalvelocity}
\end{equation}
where g is the acceleration of gravity.  However we use an upward flow
of air of velocity $v_{air}$ in the $-z$ direction.  Hence the net
downward velocity of the drop is
\begin{equation}
v_z = \frac{2 r^2 \rho g}{9\eta}-v_{air}.
\label{eq:apparentvelocity}
\end{equation}

As explained in the next section we want $v_z$ to be small hence we
set $v_{air}$ to be close to $v_{z,term}$ but slightly smaller.

\subsection{General description of experiment}

\begin{figure*}
\resizebox{1.75\columnwidth}{!}{
	\includegraphics{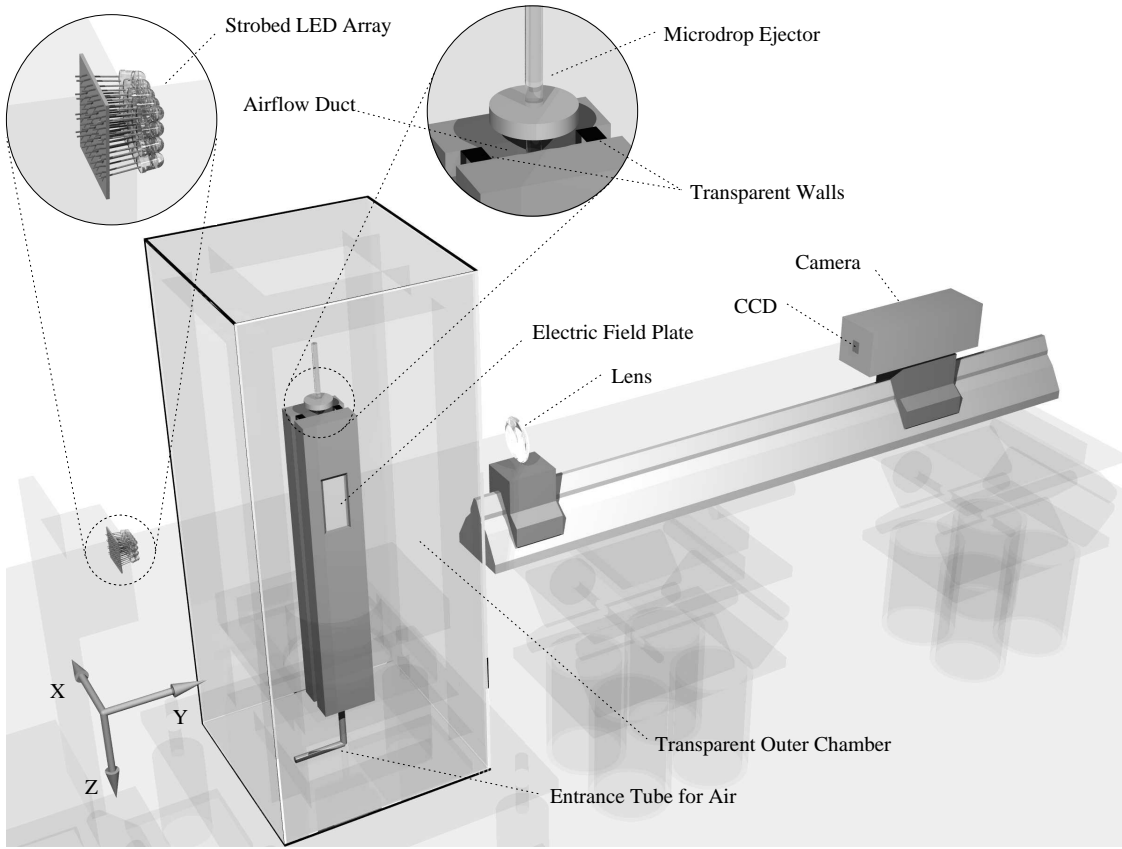}
}
\caption{Diagram of the apparatus.  Diagram is to scale, except for the
lens and CCD which are shown at 2$\times$ scale.  Support structures
are drawn transparent for clarity.}
\label{fig:schematic}
\end{figure*}

Figure~\ref{fig:schematic} is a schematic picture of the
apparatus.  Drops averaging 20.6~$\mu$m in diameter are produced at a
rate of 1~Hz using a piezoelectrically actuated
drop-on-demand microdrop ejector.  The
drops fall through the upward moving air in the measurement chamber
passing through a horizontal, uniform, alternating electric field.  In
this figure the electric field is perpendicular to the paper.  The
electric field alternates as a square wave with a frequency of 2.5~Hz
and has an amplitude of about $1.8\times 10^6$~V/m.

A rectangular measurement region 2.29~mm in the $x$ direction by 3.05~mm
in the $z$ direction is projected by a lens onto the charge-coupled
device (CCD) sensor of a
monochrome, digital video camera.
A light source consisting of a bank of light emitting diodes (LEDs)
provides 10~Hz stroboscopic
illumination.  As the motion of the drop carries it through the measurement
region, its image appears on the surface of the CCD.
Thus the camera collects 10 frames per second, the drop
appearing as a dark image on a bright background.

In addition to the $v_x$ motion there is also the $v_z$ motion,
\eqn{apparentvelocity}.  Since the camera has a field of view of
$Z=3.05$~mm in the vertical direction, the 10~Hz stroboscopic
illumination leads to acquisition of
\begin{equation}
N_{images} = 10Z/v_z = 30.5/v_z
\label{eq:nimages}
\end{equation}
images of any given drop, before the drop moves below the viewing area
of the
camera.  Here $v_z$ is in mm/s.  Hence we get a larger number of images
per drop, leading to better charge measurement precision, when $v_z$ is
small.  Of the order of $N_{images}=15$ are required.

We give an example of the importance of the upward airflow in
obtaining this many images.  Consider a typical drop of diameter
20.6~$\mu$m with a density of 0.913~g/cm$^3$.  From \eqn{terminalvelocity},
$v_{z,term} = 11.3$~mm/s.  If there were no upward airflow there would
be an average of 2.6 images per drop.  To obtain $N_{images} = 15$,
$v_z$ must be about 2.0~mm/s.  Therefore from \eqn{apparentvelocity},
$v_{air}$ must be $11.3 - 2.0 = 9.3$~mm/s.

Each image from the CCD camera is processed though a framegrabber in
a conventional desktop computer, the signal in each pixel being
recorded.  An analysis program then finds the drop images and
calculates the $x$ and $z$ coordinates of the centroid of the drop
image.  Using all the images of the drop and knowing the time spacing
of the images, $v_x$ and $v_z$ are then calculated.

\subsection{Drop generator}
\label{sec:drop_generator}

\begin{figure}
\resizebox{\columnwidth}{!}{
	\includegraphics{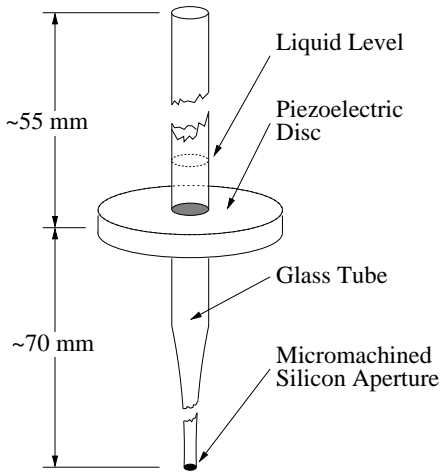}
}
\caption{Drop generator.}
\label{fig:dropper}
\end{figure}

\begin{figure}
\resizebox{0.7\columnwidth}{!}{
	\includegraphics{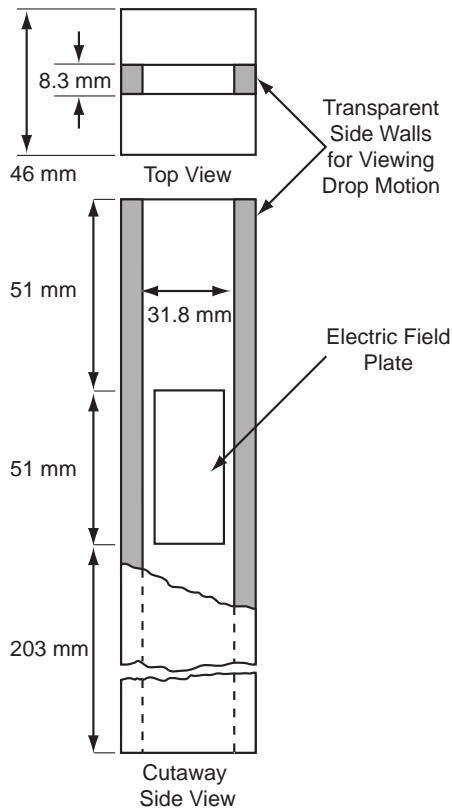}
}
\caption{Airflow tube and measurement chamber.}
\label{fig:airflow_tube}
\end{figure}

\begin{figure*}
\resizebox{1.75\columnwidth}{!}{
	\includegraphics{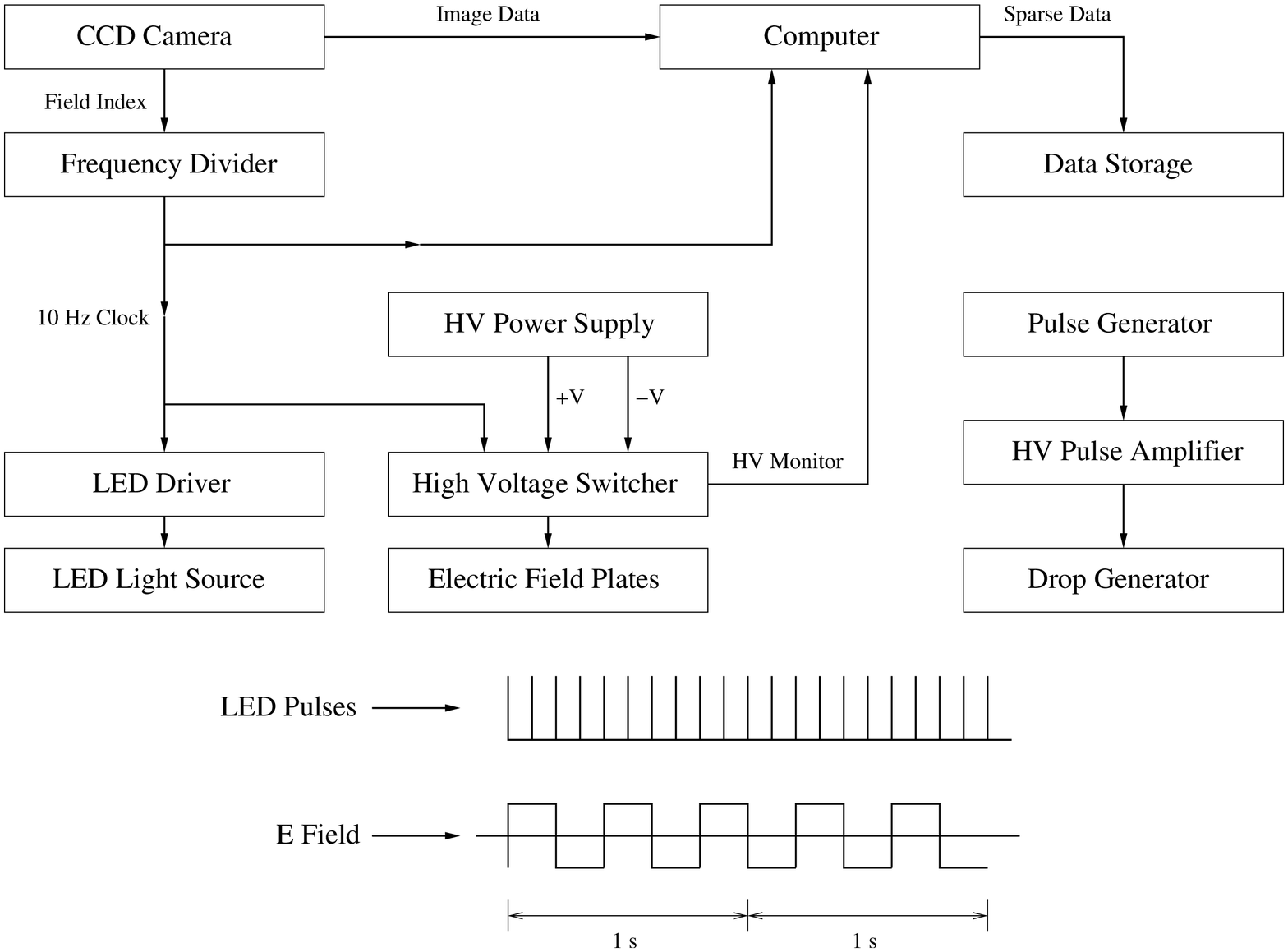}
}
\caption{Schematic of electronic system.  The LED and high voltage is
synchronized to the 10 Hz clock, while the drop generator runs asynchronously.}
\label{fig:block_diagram}
\end{figure*}

The drop generator, \fig{dropper}, is based on the general design principles
used in piezoelectrically actuated, drop-on-demand,
inkjet print heads.  Our generator is
designed for flexibility, allowing a variety of liquids to be used and
providing ease of control and maintenance.  The body of the generator is
glass so as to preserve the purity of the liquid, but the ejection aperture
at the bottom is micromachined silicon \cite{patent}.  The diameter of the
aperture sets approximately the diameter of the drop.  Upon application of a
short voltage pulse, usually 3 to 20~$\mu$s, across the surfaces of the
piezoelectric disk, the central hole in the disk contracts in diameter,
squeezing the glass tube and sending a pressure pulse down the liquid,
ejecting a jet of fluid from the aperture.
The forming of a discrete fluid drop from the high speed jet is a complex
process with the repeatability of the process and the final diameter of the
drop being highly dependent on the properties of the fluid, and on how the
fluid is driven.
For this reason, the shape and amplitude of the voltage
pulse applied to the piezoelectric disk must be specifically tuned for
stability and the desired drop size.
In addition, it is necessary to experiment with single and double pulsing,
varying both the pulse width and the separation between the pulses.

The pressure in the drop generator
is maintained slightly below atmospheric pressure by 10 to 30~mm Hg.  This
helps to retract the excess liquid outside the ejection aperture after the
drop has been produced and also prevents leaking of the liquid between pulses.

It is important that the drop generator produce drops of constant radius,
the primary reason being that the size of the drops determines the ratio
between $Q$ and the measured quantity $v_x$, from \eqn{stokes}.
A secondary reason that we did not initially appreciate has to do with
Eqs. \ref{eq:terminalvelocity} and \ref{eq:apparentvelocity}.  We set
$v_{air}$ close to $v_{z,term}$ so that $v_z$ is much smaller than
$v_{z,term}$.  Thus a small change in $r$ leads to a relatively large
change in $v_z$.  This dispersion disrupts the consistent spacing between
adjacent drops, and a decreased separation between drops is undersirable
for reasons discussed in \sect{mindist}.

With a clean and newly tuned drop generator we get remarkably uniform drop
radii, constant to about $\pm0.2\%$.  The drop generator also ejects in a
consistent downward vertical direction along the centerline of the airflow
tube.
At 1~Hz operation, the drop ejector exhibits slow
drifts in its characteristics with time scales of the order of a week.
These drifts appear as changes in drop size and as destabilization that
manifests as the appearance of satellite drops or inconsistent drop
production.  Typically, these effects can be compensated for by small
changes in the drive parameters or adjustments of the air velocity or both.
By the end of the first data set, set~1, the drop
ejector had destabilized to
the point where it had to be removed from the apparatus for cleaning and
refilling.  Similarly during the taking of data set~2, the
drop ejector and
air velocity required periodic small adjustments.  The end of set~2 was
caused by increasing instability in drop production which could not be
compensated for.  We do not know the reason for this behavior.

In our drop generator the silicone oil drops are produced with a spread of
charges, $|Q|$ ranging mostly from 0 to about 10~$e$.  A few percent of the
drops have larger $|Q|$.  As described in \sect{abs_q_criterion} we used
drops with
$|Q| < 9.5 e$ to maintain good precision in the charge measurement.  We
do not know what sets the charge distribution for a particular drop
generator.  But we have the general observation that silicone oil gives
narrow charge distributions, whereas water, mineral oil and most other
organic liquids give broad charge distributions, with $|Q|$ values as large
as several 1000~$e$ or even larger.

\subsection{Optical system}
\label{sec:optical_system}

Referring to \fig{schematic}, the stroboscopic light source consists of a
rectangular bank of 20~LED's emitting at 660~nm.
The pulse length was about 100~$\mu$s.
The lens, a 135~mm focal length, $f$/11, photographic enlarging lens, images
the measurement region
onto the face of the CCD camera with a magnification of 2.1.

The rectangular active image area of the CCD camera \cite{camera} is 4.8~mm
in the horizontal, that is, $x$, direction, and 6.4~mm in the vertical,
that is,
$z$, direction.  Hence the viewing area in
physical space is 2.29~mm horizontally
by 3.05~mm vertically.  We remind the reader that the electric field is
horizontal.  The active imaging area is an array of 240 horizontal picture
elements (pixels) and 736 vertical picture elements (pixels).  We chose this
orientation of the array to maximize the vertical distance, maximizing the
number of images per drop.

Given the magnification and pixel density of the CCD, one would expect from
geometric optics that the shadow of a 20~$\mu$m diameter drop would cover
2 pixels horizontally and 5 pixels vertically.  The actual observed shadow
typically covered 3 pixels horizontally and 7 pixels vertically, and had an
intensity variation that was approximately a two dimensional Gaussian.  This
can be quantitatively described as the convolution of the simple shadow
predicted by geometric optics with a point spread function that is a result
of the diffractive effects due to the finite aperture of the lens.  We do not
and should not observe diffractive effects caused by the small size of the
drops.

\subsection{Airflow tube and measurement chamber}

Figure~\ref{fig:airflow_tube} shows a slightly simplified,
dimensioned drawing of the airflow tube
and the measurement chamber.  A rectangular duct contains the
upward flowing air.  It is 8.3~mm wide in the direction of the electric field
and 31.8~mm wide in the direction perpendicular to the electric field.  The
field plates that define the measurement chamber are 51~mm high and
28.6~mm wide.  The inner surfaces of the plates are machined flat
and are in the same plane as the inner surfaces of the walls of the airflow
tube.  The optic axis of the optical system passes through the transparent
side walls of the airflow tube.

The air velocity is sufficiently small, with a Reynolds number on the order
of $R_e=50$ so that the flow is laminar.  The 203~mm length of air flow tube
between the measurement region and the air inlet allows the air to settle
into its equilibrium flow pattern.  At equilibrium, the velocity profile of
the air is approximately parabolic across the narrow direction of the
channel ($x$~axis).  Across the long axis, the flow has a roughly constant
central region and falls to zero at the boundaries \cite{langlois}.

\subsection{Electronics}

All the electronics of the apparatus, \fig{block_diagram}, are hard-wired to
give reliable timing, independent of the operation of the computer.  A 30~Hz
handshaking signal from the CCD is divided down to provide a 10~Hz clock that
synchronizes the LED strobe, the electric field switcher and the computer
image acquisition.  The switching of the electric field, which is driven by
the clock signal divided by 4, operates at 2.5~Hz.  This results in a cycle
where two images are acquired with the electric field in one direction, and
then two images with the electric field in the other direction.  These
relationships between the signals is depicted in the timing diagram of
\fig{block_diagram}.  The drop generator is driven asynchronously at 1~Hz.

\subsection{Data acquisition and storage}

Data acquisition was performed by a single desktop computer running Linux.
The computer was equipped with two special components: a digital
framegrabber that allowed the capture of image data from the camera and a
general purpose input/output interface board with digital I/O and A/D
conversion capability.  The additional inputs allowed the computer to
monitor the state of the experiment as well as a variety of environmental
variables.

All software was custom written in C.  Hardware dependent code was
encapsulated into drivers at the kernel level, which allowed a guarantee
of synchronization of the software with apparatus by a combination of
hardware and software buffering of the data.

The overall strategy was to acquire data from the apparatus and write them
to files in \emph{raw} form for later processing off line.  Recall that each
image frame contains $736 \times 240$ pixels, each digitized to 8~bit
accuracy.  Therefore acquiring data at 10~Hz produces a data rate from the
CCD camera of about 2~MB/s, much too large to be stored.  Since an image
contains just a few drops, most of the pixels in an image have just the
background signal, which allows the information to be stored using sparse
storage.  As mentioned in \sect{optical_system},
the typical image of drop extends over
an area of 3~pixels by 7~pixels.  In each frame the regions containing drops
were isolated using a thresholding operation.  The position of each drop was
then measured using a simple center of mass algorithm, and for each drop only
a surrounding region containing 13 horizontal pixels by 21 vertical pixels of
the image is written to the output file.

\subsection{Data collection}

The search was carried out in two sets described in
Table~\ref{tab:data_collection}.
At the beginning of the Set~1 the drop ejector was operated at 0.5~Hz, then
at 1~Hz for the remainder of Set~1 and all of Set~2.

\begin{table}[htb]
\caption{Data collection}
\label{tab:data_collection}
\vspace*{10 pt}
\begin{tabular}{cccc} \hline\hline
Data set & Weeks & Number of drops & Total mass (mg) \\ \hline
1 & 13 &   3 377 477  & 12.1 \\
2 & 17 & 13 430 167 &  58.0 \\ \hline\hline
\end{tabular}
\end{table}

\section{CALIBRATION, ERRORS, AND MEASUREMENT PRECISION}
\label{sec:calibration_and_errors}

\subsection{Electric field and drop radius}
\label{sec:field_and_radius}

Rewrite \eqn{stokes} in the form 
\begin{equation}
v_x = \left(\frac{Q}{6\pi \eta }\right)\left(\frac{E}{r}\right)
\label{eq:stokes2}
\end{equation}

Consider nonfractional values of $Q=ne$,  $n=0,\pm1,\pm2,\ldots$.  Then,
as shown in \sect{results}, the measured values of $v_x$ sharply peak at
$n=0,\pm1,\pm2,\ldots$.
From a fit to the center of these peaks, \eqn{stokes2} gives the fitted
$E/r$ ratio.

The electric field strength, $E$, is calculated from the measured voltage
across the electric field plates, known to 3\%, and the plate separation,
8.25$\pm$0.01~mm.  The plates are parallel to within 0.1~mrad.  Inserting
the calculated value of $E$ into the fitted $E/r$ ratio, we obtain the drop
radius $r$. 

We have two additional checks of the drop radius, one from the measurement
of the error caused by the Brownian motion, \sect{brownian_centroiding},
and the other from the
measurement of the net downward velocity of the drops, $v_z$ in
\eqn{apparentvelocity}.

The drop radius depends to a moderate extent upon the size and shape of
the voltage pulse applied to the drop generator and to a slight extent
upon the age and history of the drop generator.  However over periods of
hours the average drop radius could be taken to be constant, with
fluctuations of $\pm0.2\%$ for individual drops.  Since the data were
analyzed in one-hour-long blocks, the average $E/r$ ratio for any given
block was known to much better accuracy than this.

\subsection{Brownian motion and drop position measurement errors}
\label{sec:brownian_centroiding}

The precision of the determination of the drop charge depends upon the
precision of the measurement of $v_x$.  Consider the sequence of position
measurements $x_i$ of the trajectory of a drop.  For two consecutive frames,
$j$ and $j-1$, the velocity measurement $v_{x,j}$ is given by
\begin{equation}
v_{x,j} \equiv \frac{x_j-x_{j-1}}{\Delta t}.
\label{eq:vx}
\end{equation}
Here,
$\Delta t$ is the time between successive frames, 0.1~s in our case.  Since
$\Delta t$ is known with very good precision, the error in measuring $v_x$
comes from the error in determining the $x_i$ of the drop centers,
and from Brownian motion.
Take the error in centroiding to be normally
distributed with a standard deviation of $\sigma_c$.

During the time $\Delta t$ between any two
successive measurements of the $x_i$
positions, Brownian motion adds a random
contribution with standard deviation given by
\begin{equation}
\sigma_b=\sqrt{\frac{kT\Delta t}{3 \pi \eta r}}.
\label{eq:sigb}
\end{equation}
Here $k$ is the Boltzmann constant, $T$ is the absolute temperature, $\eta$
is the viscosity of air and $r$ is the drop radius.

The trajectory of uncharged drops, which have no contribution to their
trajectories due to the electric field, can thus be written as
\begin{eqnarray}
x_j = x_0 + \sigma_{c,j} + \sum_{i=1}^j \sigma_{b,i} \\
j=1,2,\ldots,N_{images}\nonumber
\end{eqnarray}
with $x_0$ set by the initial position of the drop, and
where the $\sigma_{c,i}$ ($\sigma_{b,i}$)
are normally distributed with a std. dev. of $\sigma_c$ ($\sigma_b$).
The analysis for charged drops is similar if the effect of the electric
field is first subtracted from the observed data points.
It then follows that
\begin{equation}
v_{x,j}\Delta t = \sigma_{b,j} + \sigma_{c,j} - \sigma_{c,j-1}
\end{equation}
and
\begin{equation}
\label{eq:errors}
\left< v_{x,j} v_{x,k} \right>\Delta t^2 = \left\{
	\begin{array}{cl}
		2\sigma_c^2 + \sigma_b^2, &j=k \\
		-\sigma_c^2, &|j-k|=1 \\
		0, &\mbox{otherwise.}
	\end{array} \right.
\end{equation}
Therefore the total error on any given velocity measurement, $\sigma_v$, is
given by $\sigma_v^2=2\sigma_c^2+\sigma_b^2$, and
the centroiding error introduces an anticorrelation with magnitude
$-\sigma_c^2$ in two consecutive velocity measurements, due to the
shared position measurement.  We use the concept summarized in
\eqn{errors} and the observed distributions of the $v_{x,i}$
(after removal of the contribution due to the alternating electric
field) to separate $\sigma_c$ from $\sigma_b$.

We find that averaged over this experiment
\begin{equation}
\sigma_c =0.31\mbox{ $\mu$m} \mbox{, } \sigma_b =0.47\mbox{ $\mu$m},
\label{eq:sig2}
\end{equation}
in the measurement region.  Compared to the size of an individual pixel
on the CCD, the centroiding error is small, approximately 1/30 of a pixel.
The value of $\sigma_b$ obtained provides an independent check on the
size of the drops, and is consistent with the size determined from the
terminal velocity and the electric field drift velocity.

Equation~\ref{eq:sig2} shows that the Brownian motion error, $\sigma_b$, is
about the same magnitude as the error involved in finding the drop position,
$\sigma_c$.  Therefore substantially reducing $\sigma_c$ through the use of
smaller pixels will not by itself substantially reduce the error on the
charge measurement, since
the Brownian motion error can only be reduced by increasing $N_{images}$.

The final charge measurement of a drop is made using a single, detailed
best fit
to the entire observed trajectory of the drop, and the final error on the
charge measurement $\sigma_q$ is a result of propagating the errors
$\sigma_c$ and $\sigma_b$ through this calculation.

\subsection{Other sources of errors}
\label{sec:other_errors}

We looked for other sources of errors, but all are negligible compared to
those in \eqn{errors}.  When we developed the upward air flow method we thought
about the possibility that there might be some horizontal air velocity,
$v_{x,air}$ in the measurement chamber, contributing an error to
$\sigma_v$ of order $v_{x,air}\times\Delta t$.  By studying a
large amount of data we found that the distribution of
$v_{x,air}\times\Delta t$ had an rms value of 100~nm, and was a fixed
property of the measurement region.  For comparison $v_{e}\times\Delta t$
was of the order of 8~$\mu$m. These irregularities in $v_{x,air}$ are
probably due to residual surface imperfections in the electric field plates.
Since the irregularities are constant over long periods of time, they can
be accurately measured and corrected for.  For this analysis, that was not
necessary.

Another possible source of error would be a nonuniformity in the electric
field in the measurement region giving a horizontal gradient, $dE/dx$.
This would produce a horizontal force on the drop's induced electric dipole
moment.  This dipole force acts in addition to the $QE$ force.  We found
such a dipole force to be negligible compared to the $QE$ force.

A small, vertical deceleration of the drops as they fall through the
measurement chamber was observed.  This amounted to a change of 30~$\mu$m/s
in the apparent terminal velocity of the drops as they fell through the
measurement region, or a systematic uncertainty in the radius of the
drop of the order of 0.3\%.
We
believe that the deceleration is due to the evaporation of the drop as it
falls.
The magnitude of this effect is small enough such that it can be neglected
in the calculation of $v_x$.  As a side note, any systematic uncertainties
in the radius of the drops are absorbed by the calibration process
described in \sect{field_and_radius}, and do not affect the final charge
measurement.

Similarly, other possible sources of measurement error such as apparatus
vibrations, optical distortions and CCD array distortions,
and patch nonuniformities on the electric field plates, were negligible.

\section{DATA ANALYSIS AND RESULTS}
\label{sec:analysis_and_results}

\subsection{Drop selection criteria}

In this section we use $q=Q/e$, a measure of the drop charge in units of the
electron charge.  We required that all drops used in the analysis meet the
criteria in Table~\ref{tab:criteria}.  The criteria are designed to maintain
a charge
measurement accuracy of approximately 0.03~$e$ and to reject irregular drops
caused by inconsistent operation of the drop generator.

\subsubsection{$q < 9.5$ criterion}
\label{sec:abs_q_criterion}
 
For any given drop there is an uncertainty in the radius of approximately
0.2\% which contributes to the relative error on $q$. The \emph{absolute}
error on $q$ thus increases linearly with $q$. Since the absolute error on
$q$ must be kept to the order of 0.03~$e$, restricting the data sample to
drops with $q<9.5$ keeps this contribution to less than 0.02~$e$. The
overall charge distribution is such that only a few percent of the drops
have $q$ values outside this range.

\subsubsection{$\sigma_q < 0.03$ criterion}

Primarily, this criterion is a measure of $N_{images}$ of the drop.
Brownian motion and centroiding accuracy, characterized by $\sigma_c$ and
$\sigma_b$ as described in \sect{brownian_centroiding}, limit the
accuracy of the charge measurement, $\sigma_q$.  For any given drop,
the number of position measurements, $N_{images}$, and the state of the
electric field during those measurements, in addition to $\sigma_b$ and
$\sigma_c$, determines this accuracy.
As noted earlier, $N_{images}$ was of the order of 15.  If a drop has an
exceptionally large radius or is falling too far from the centerline of the
airflow tube, $v_z$ will be too large and $N_{images}$ will be too small.

\subsubsection{$\chi^2$ criterion}

As mentioned earlier, the final calculation of the charge on a drop is done
using a fit to the trajectory of the drop.  It was required that the $\chi^2$
probability of the fit to the drop's trajectory be better than $10^{-3}$.
This rejects a large class of rare artifacts based on the statistical
likelihood that the observed deviations from the fitted trajectory could be
attributed to the Brownian motion and centroiding errors.  For example, a
drop would be rejected if it had an anomalous trajectory due to vibrations
in the apparatus or due to its charge having been changed by collision with
an ion during measurement.

\subsubsection{$v_z$ criterion}

The net downward velocity of the drop, $v_z$, depends upon the drop radius
and the upward air velocity, $v_{air}$, \eqn{apparentvelocity}.  This
criterion insures consistent drop radii within the hour long data blocks
by requiring
\begin{equation}
|v_{z,drop}-v_{z,block}| < 0.124\mbox{ mm/s}
\label{eq:vzcut}
\end{equation}
where $v_z$ is the measured value for one drop and $v_{z,block}$ is the
average value of $v_z$ for all the drops in the one hour data block.
Using Eqs.~\ref{eq:terminalvelocity} and \ref{eq:apparentvelocity},
taking $v_{air}$ as fixed and using an average value
for $r$ of 10~$\mu$m, this eliminates any drops with $r$ different from the
nominal value by more than about $\pm0.5\%$.

Recall that the air velocity is approximately parabolic and that close to
the centerline it is given by 
\begin{equation}
v_{air}(x) =v_{air,0}\left[1 -(x/x_{w})^2\right]
\label{eq:parabolicairflow}
\end{equation}
where $x$ is the distance along the $x$~axis from the centerline of the
airflow tube, $x_w = 4.15$~mm is the distance to the wall of the tube,
and $v_{air,0}$ is the air velocity along the centerline. Therefore this
criterion indirectly restricts how far the drop can be from the centerline. 

\subsubsection{$x$ deviation criterion}

This criterion 
\begin{equation}
|x-x_{block}| < 0.19 \mbox{ mm}
\label{eq:avgxcut}
\end{equation}
provides a direct constraint on how far a drop may deviate from the centerline
in the $x$ direction.  Here $x_{block}$ is the average value of $x$ for all
the drops in the one hour data block. The purpose of this criterion is to
eliminate drops that were produced irregularly.  The 0.19~mm upper limit in
\eqn{avgxcut} was determined by examining the distribution of $x$ positions
of drops produced during normal operation of the drop generation and setting
the upper limit to eliminate the tails.

\subsubsection{Minimum distance $R$ between any two drops criterion}
\label{sec:mindist}

\begin{figure}
\resizebox{0.7\columnwidth}{!}{
	\includegraphics{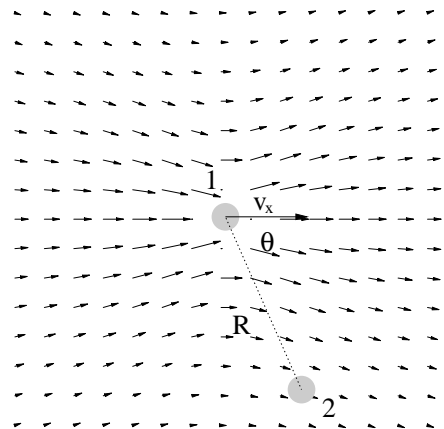}
}
\caption{The viscous coupling between a moving drop~1 on a neighboring drop~2
in still air.  The small arrows show the vector velocity of the disturbed
air.  Note that there is a slight disturbance at the position of drop~2
that will affect the trajectory of drop~2.}
\label{fig:stokes_interaction}
\end{figure}

The drops interact with one another through their
induced electric dipole moments
and viscous coupling through the air.
Consider two drops of radius $r$, drop~1 moving with a velocity $v_x$
due to the force of
the electric field on its charge, as shown in \fig{stokes_interaction}.
This motion will move the surrounding air.
At the position of drop~2, the velocity of the air in the $x$ direction,
$V_{x, \mbox{disturbed air}}$, is given by
\begin{equation}
V_{x,\mbox{disturbed air}}=\frac{3}{4}\frac{v_xr}{R}(1+\mbox{cos}^2\theta).
\label{eq:stokesflow}
\end{equation}

Since drop~2 sits in this disturbed air, its $v_x$ due to the force of the
electric field on its charge will have superimposed upon it
$V_{x, \mbox{disturbed air}}$.  This will distort the charge measurement.
Therefore $V_{x, \mbox{disturbed air}}$ must be kept small by keeping $R$,
the distance between the drops, much larger than $r$, the radius of the drop.
A large separation also serves to minimize the interaction between the
induced electric dipole moments of the drops, which increases as the inverse
fourth power of the separation.
We require
\begin{equation}
R > 0.62  \mbox{ mm}
\label{eq:mindistcut}
\end{equation}
separation between any two drops, which limits these forces to a small
fraction of $QE$.

\subsubsection{Summary and magnitude of drop selection criteria}

Table~\ref{tab:criteria} gives the percent of drops removed by each
criterion averaged over
each of the two data sets.  The total percent of drops removed is also
given.
Since the same drop may be removed by several criteria, the total percent
removed is \emph{not} the sum of the percent removed by the individual
criteria.

\begin{table}[tb]
\caption{Drop selection criteria.  The entries are the percent removed by
each criterion separately.  The bottom row gives the total percent of drops
removed by all criteria.  Since the same drop may be removed by several
criteria, the total percent removed is \emph{not} the sum of the percent
removed by the individual criteria.}
\label{tab:criteria}
\vspace*{10pt}
\begin{tabular}{crr} \hline\hline
Criterion          & \hspace{0.2in}Set~1  & \hspace{0.2in}Set~2 \\ \hline
$q$                &  0.4   &  0.7  \\
$\sigma_q$         &  6.0   &  0.3  \\
$\chi^2$           &  4.7   &  2.4  \\
$v_z$              &  4.6   &  1.2  \\
$x$                &  9.4   &  5.1  \\
$R$                & 12.5   &  3.8  \\
Total              & 22.3   &  8.7  \\ \hline\hline 
\end{tabular}
\end{table}

\subsection{Results}
\label{sec:results}

\begin{figure}
\resizebox{\columnwidth}{!}{
	\includegraphics{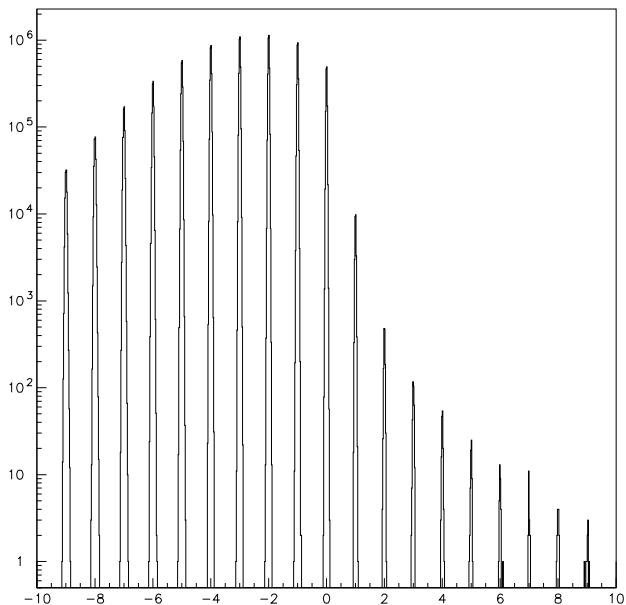}
}
\caption{The $q$ charge distribution in units of $e$.}
\label{fig:q_distribution}
\end{figure}

After the application of these criteria we had a final data sample of
$1.7\times10^7$ drops of average diameter 20.6~$\mu$m.  The total mass of
the sample was 70.1~mg.  Figure~\ref{fig:q_distribution} shows the charge
distribution in units of $e$.  (The asymmetry of  the charge distribution is
a property of the drop generator as discussed in \sect{drop_generator}.)
We see sharp
peaks at integer numbers of charges and no drops further than 0.15~$e$
from the nearest integer.  We emphasize that there is no background
subtraction here, this is all the data after the application of the criteria
previously discussed. 

\begin{figure}
\resizebox{\columnwidth}{!}{
	\includegraphics{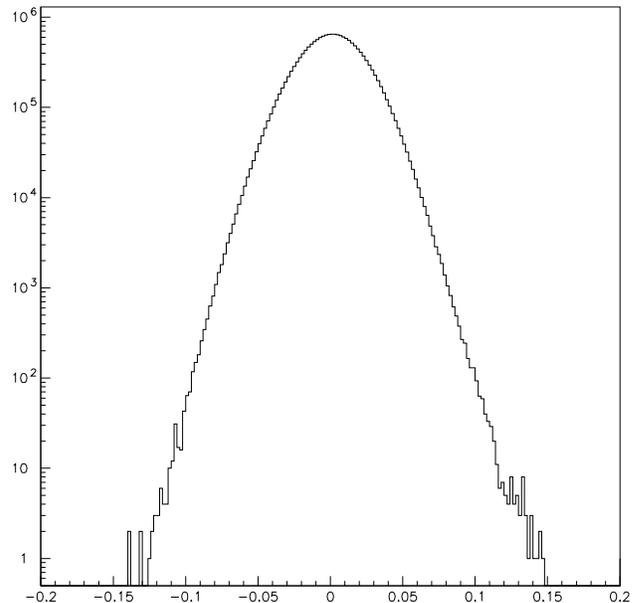}
}
\caption{The $q_c$ charge distribution in units of e.}
\label{fig:q_residual}
\end{figure}

To show the shape of the peaks at integer values of $q$ we superimpose them
in \fig{q_residual} using the charge distribution, $q_c$, defined by
$q_c=q-N_c$ where $N_c$ is the signed integer closest to $q$.  The peaks
have a Gaussian distribution with a standard deviation of 0.021~$e$.
\emph{The absence of non-Gaussian tails is what allows this search method
to be so powerful.}

\begin{figure}
\resizebox{\columnwidth}{!}{
	\includegraphics{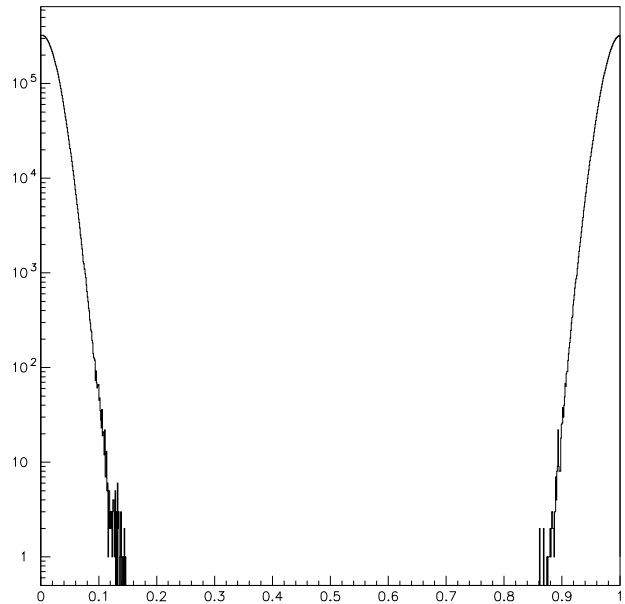}
}
\caption{The $q_c$ residual charge distribution in units of e.}
\label{fig:q_exclusion}
\end{figure}

In \fig{q_exclusion} we superimpose the valleys between the peaks using the
residual charge distribution, $q_r$, defined by $q_r=q-N_l$ where $N_l$ is
the largest integer less than $q$.  We did not find any drops with residual
charge between 0.15~$e$ and 0.85~$e$.  In this residual charge range there
are fewer
than $1.17\times10^{-22}$ fractional charge particles per nucleon in silicone
oil with 95\% confidence.

Thus this 70.1~mg search did not confirm the one unusual aspect of our
previous 17~mg search, Halyo \etal \cite{halyo}, where we found 1~drop with
a $q_r$ of about 0.29~$e$.  No such charge was found in this search.  While it
is of course still possible that the fractional charge found in the 17~mg
experiment was real, we are inclined to believe that the 17~mg experiment had a
very small background that has been eliminated by the improved method of this
experiment.

\section{CONCLUSIONS}
\label{sec:conclusions}

\subsection{Comparison with other fractional charge searches in bulk material}

In Table~\ref{tab:searches} we compare this search with previous, larger
sample, searches for
fractional charge particles in bulk matter.  No evidence for fractional charge
particles was found in the searches by Marinelli \etal \cite{mari}, Smith
\etal \cite{smi3}, and Jones \etal \cite{jon2},  similar to the null result in
the present search.

In their superconducting levitometer search in niobium, LaRue \etal \cite{lar}
claimed to have fractional charge particles with $e/3$ and $2e/3$.  But
Smith \etal \cite{smi3} who also searched in niobium using a ferromagnetic
levitometer method did not find any such evidence in a four times larger
sample.  At present the results of LaRue \etal \cite{lar} are not understood
and are generally not accepted.

Our search is by far the largest to date and has the smallest upper limit
of any search on the concentration of fractional charge particles in bulk
matter.  But it is important not to generalize our limit to other kinds of
bulk matter for several reasons.  First, we do not know what happens to
fractional charge particles that are in natural matter when that matter is
processed.  Note that except for the search in meteoritic material by Jones
\etal \cite{jon2}, all the material in Table~\ref{tab:searches} is processed.

Second, if we assume the existence of stable, fractional charge particles,
we do not know what natural materials are most likely to have a detectable
concentration.  Our own thoughts are that the most promising natural
material is that found in carbonaceous chondrite asteroids,
since they are representative of the
primordial composition of the solar system, having not undergone any
geochemical or biochemical processes.
Hence similar to the motivation of Jones
\etal \cite{jon2}, our next search will be in meteoritic material from an
asteroid.

\begin{table}
\caption{Searches for fractional charge particles in ordinary matter.  All
experimenters reported null results except LaRue \etal \cite{lar}.  There
are $6.4\times10^{20}$ nucleons in a milligram.}
\label{tab:searches}
\vspace*{10pt}
\begin{tabular}{lllc} \hline\hline
Method & Experiment & Material & Mass(mg)\\ \hline
levitometer & LaRue \etal \cite{lar}        & niobium        &  1.1 \\
levitometer & Marinelli \etal \cite{mari}   & iron           &  3.7 \\
levitometer & Smith \etal \cite{smi3}       & niobium        &  4.9 \\
levitometer & Jones \etal \cite{jon2}       & meteorite      &  2.8 \\ 
liquid drop & Halyo \etal \cite{halyo}      & silicone oil   & 17.4 \\
liquid drop & this search                  & silicone oil   & 70.1 \\
\hline\hline
\end{tabular}
\end{table}

\subsection{Remarks on further use of this new method}

The purpose of the new method \cite{loomba} used in this experiment was to
allow large drops to be used compared to the classical method, thus
increasing the rate at which we could search through a sample and also
enabling the use of suspensions of more interesting materials.
We have
succeeded in doing this, using drops of about 20~$\mu$m diameter compared
to the approximately 10~$\mu$m diameter used in Halyo \etal \cite{halyo}.
In the Appendix we discuss further increasing the search rate by using
still larger drops and by using multiple columns of drops to increase the
total rate of drop production.  We find that with this new method the mass
per second search rate can be further increased by a factor of
the order of 10, but probably not by a factor of a 100.

%PRD_ONLY\acknowledgments

%PRD_ONLYThis work was supported by Department of Energy contract DE-AC03-76SF00515.

\appendix*
\section{INCREASING THE SEARCH RATE}

Three are several ways in which the mass per second search rate can be
increased in this experimental method.

\subsection{Use of larger drops}

The first way to increase the search rate is to use larger drops.
Maintenance of the precision of the charge measurement requires that
$N_{images}$ increase in proportion to the drop radius.  An increase in
$N_{images}$ can be accomplished by some combination of a decrease in $v_z$
and an increase in the vertical length $Z$, \eqn{nimages}.
However a significant
decrease in $v_z$ requires too fine a balance between $v_{air}$ and $r^2$,
\eqn{apparentvelocity}.  If we keep $v_z$ constant, an increase of $Z$ can
be attained by an increase in the number of vertical direction pixels in the
CCD array of the camera.  Existing CCD cameras with 10~Hz frame reading
rates have twice the 736 vertical pixels used in the present camera and
larger arrays will probably be available in the future.  Therefore based on
this consideration alone, drop diameters of several times 20~$\mu$m are
feasible.

However, there are two problems that must be considered for drop diameters
larger than 30 to 40~$\mu$m.  The dipole force on a drop in a non-uniform
electric field is proportional to the third power of the drop diameter.  This
force was negligible in this experiment, \sect{other_errors},
but would have to be
considered for much larger drops.  The other problem is that the maintenance
of a small and constant $v_z$, \eqn{apparentvelocity}, requires $v_{air}$ to
increase as the square of the drop diameter, possibly leading to non-laminar
flow.  Therefore without more design and experimental studies, our
conservative conclusion is that the drop diameter is limited to about
30~$\mu$m.  This would lead to an increase of the mass search rate by a
factor of 3 compared to the 20~$\mu$m drops used in this experiment.

\subsection{Increase of drop production rate per column of drops}

Let the drop production rate for a column of drops be $n$ per second.
Then the vertical separation between drops in a column is $R=v_{air}/n$.
The criteria in
\sect{mindist} require $R>0.62$~mm.  Using $v_{air}=2.0$~mm/s,
this gives an upper limit on $n$ of about 3~Hz.  However our experience in
this experiment, \sect{drop_generator},
strongly suggests that a maximum 1~Hz rate
is conservative practice, because of irregularities in drop production.

\subsection{Increase in the number of drop columns}

In this experiment we used one column of drops, however the extension to
many columns of drops is straightforward.  Of course the horizontal
separation between adjacent columns must meet the requirements of
\eqn{mindistcut}, a nominal separation of 1~mm is useful for design
purposes.  The use of multiple columns requires two changes in the
experimental design, namely the number of pixels in the horizontal direction in
the CCD array must be increased and the space between the electric field
plates must be increased.  The latter requirement means the alternating
potential
applied across the plates must also be increased to keep the electric
field constant. 

Existing 10~Hz frame rate CCD cameras limit the number of columns to three
but improvements in these cameras would probably allow five columns.  The
corresponding increase in the electric plate spacing and the potential
difference is straightforward.

\subsection{Correction for drop to drop interactions}

It is clear that the primary constraint limiting the density of drops
achievable in the measurement chamber is that of \eqn{mindistcut}.  To
reiterate, interaction between the drops due to their induced
electric dipole moment and viscous coupling requires that there be
a minimum separation allowable between drops.  In the limit that
these interactions are small, both of these effects can be calculated
from first principles, for example as in \eqn{stokesflow}.
In principle then, it should be possible to subtract the effect of
these perturbing forces from the measured
trajectory of each drop.  Given
this, it would be possible to relax the constraint on $R$.  Since
this possibility requires further study, it is not clear to what extent
$R$ can be reduced and throughput increased.

\subsection{Summary}

Putting these estimates together we can see how to achieve
an improvement on the order of 10 times
the present mass per second search rate using existing CCD cameras.  Future
cameras will probably allow a factor of 15 improvement.


\begin{thebibliography}{000}
\bibitem{camera}
The CCD camera, a Cohu 4110, uses the frame transfer method and has digital output.
\end{thebibliography}
\end{document}